\documentclass[sn-vancouver,Numbered]{sn-jnl}

\usepackage{graphicx}%
\usepackage{multirow}%
\usepackage{amsmath,amssymb,amsfonts}%
\usepackage{amsthm}%
\usepackage{mathrsfs}%
\usepackage[title]{appendix}%
\usepackage{xcolor}%
\usepackage{textcomp}%
\usepackage{manyfoot}%
\usepackage{booktabs}%
\usepackage{algorithm}%
\usepackage{algorithmicx}%
\usepackage{algpseudocode}%
\usepackage{listings}%

\theoremstyle{thmstyleone}%
\theoremstyle{thmstyletwo}%

\theoremstyle{thmstylethree}%

\begin{document}

\title[Article Title]{\textbf{Score-based Generative Diffusion Models to Synthesize Full-dose FDG Brain PET from MRI in Epilepsy Patients}
}


\author[1]{Jiaqi Wu}

\author[2]{Jiahong Ouyang}

\author[3]{Farshad Moradi}

\author[4]{Mohammad Mehdi Khalighi}

\author[5]{Greg Zaharchuk}

\affil*[1,2,3,4,5]{Department of Radiology, Stanford University}


\abstract{\textbf{Purpose:} Fluorodeoxyglucose (FDG) PET is one of the most common applications for simultaneous PET/MRI, given the need to image both the brain structure and its metabolism with great detail, but is suboptimal due to the radiation dose in this young population. Little work has been done synthesizing diagnostic quality PET images from MRI data or MRI data with ultralow-dose PET using advanced generative AI methods, such as diffusion models, with attention to clinical evaluations tailored for the epilepsy population.

\textbf{Methods:} We compared the performance of diffusion- and non-diffusion-based deep learning models for the MRI-to-PET image translation task for epilepsy imaging using simultaneous PET/MRI in 52 subjects (40 train/2 validate/10 hold-out test). We tested three different models: 2 score-based generative diffusion models (SGM-Karras Diffusion [SGM-KD] and SGM-variance preserving [SGM-VP]) and a Transformer-Unet. We report results on clinically relevant metrics, including congruency measures (Congruence Index and Congruency Mean Absolute Error) that assess hemispheric metabolic asymmetry, which is a key part of the clinical analysis of these images. We compared the model performance using different inputs such as T1-weighted (T1w), T2-FLAIR (T2F), and 1\% ultralow-dose PET images to evaluate the effect and necessity of each imaging contrast.

\textbf{Results:} The SGM-KD produced the best qualitative and quantitative results when synthesizing PET purely from T1w and T2-FLAIR images with the least mean absolute error in whole-brain specific uptake value ratio (SUVR) and highest intraclass correlation coefficient. When 1\% low-dose PET images are included in the inputs, all models improve significantly and are interchangeable for quantitative performance and visual quality.

\textbf{Conclusion:} SGMs holds great potential for pure MRI-to-PET translation, while all 3 model types can synthesize full-dose FDG-PET accurately using MRI and ultra-lowdose PET. This suggests that deep learning diffusion models could reduce or eliminate radiation dose for patients being evaluated for epilepsy. 

}




\maketitle

\section{Introduction}\label{sec1}
Epilepsy, a prevalent neurological disorder characterized by recurrent seizures, necessitates precise diagnostic approaches to ensure effective management. \textsuperscript{18}F-fluoro-2-deoxyglucose
(\textsuperscript{18}F-FDG) Positron Emission Tomography (PET) imaging of brain glucose metabolism is a well-established and widely available technique for localizing epileptogenic foci within the brain. PET provides functional information by detecting metabolic changes in the brain, which are crucial for localizing epileptogenic zones (EZs), especially in cases where MRI findings are inconclusive. Meanwhile, Magnetic Resonance Imaging (MRI) is an important diagnostic tool, offering detailed anatomical images and enabling the identification of structural abnormalities such as mesial temporal sclerosis and focal cortical dysplasias that may underlie epileptic activity \cite{neuroimaging_epilepsy}. Research has shown that a multi-modal platform approach integrating the functional imaging of PET with the morphologic information from MRI in presurgical evaluation of epilepsy can greatly improve outcomes \cite{role_of_spec_pet}. One study also suggests that diagnostic accuracy for EZ detection in focal epilepsy could be higher in FDG-PET/MRI compared with FDG-PET/CT \cite{Kikuchi2021}.

However, obtaining PET images poses several challenges. PET scanners are far less accessible than MRI scanners, particularly in middle- and low-income countries. Even in high-income countries, accessing PET imaging often involves long-distance travel and significant expenses \cite{Gallach2020}. Moreover, FDG-PET involves the injection of a radioactive tracer into the body, with the added absolute risk of subsequent malignancy from FDG-PET/CT estimated at 0.5\%–0.6\% \cite{pet_cancer}. In contrast, MRI is a widely available imaging technique that does not involve ionizing radiation. Therefore, exploring the feasibility of synthesizing FDG-PET images from multi-contrast MRI images or MRI plus lower dose FDG-PET is a worthwhile endeavor.

Meanwhile, artificial intelligence has shown impressive performance for medical image translation. U-net models using nested convolution neural networks (CNNs) represent some of the earliest work in this area \cite{XIANG2017406}. More recently, Generative Adversarial Networks (GANs) have been added to the basic U-Net structure to improve performance. For instance, MedGAN \cite{medgan} was developed for conversion of PET to CT images. Swin-Transformer based GAN \cite{swin} was used for multi-modal MRI translation, while Transformer-U-Net models have shown promise for multi-modal MRI-to-PET translation and medical image segmentation \cite{transunet, transunet2}. The advances in diffusion-based models such as Diffusing Denoising Probablistic Models (DDPM) have been explored by prior work \cite{fastddpm, ct_mri_ddpm, Singh_2024, Gong2024} in multi-contrast MRI sequence-to-sequence translation, CT-to-MRI translation, and PET reconstruction, demonstrating better performance than CNN and GAN models.

Although there is abundant literature leveraging 2D models for 3D volume in the domain of medical imaging, MRI-to-PET translation using generative models in the context of epilepsy remain largely under-explored, in part due to the lack of large public datasets. Non-epilepsy public MRI and PET datasets also suffer from the lack of access to list-mode raw PET data to allow for realistic synthesis of low-dose images. Additionally, the current evaluation metrics such as structural similarity (SSIM), peak signal-to-noise ratio (PSNR), root mean squared error (RMSE), frequently used in natural image assessments, are not uniformly applicable to medical images. In the context of PET synthesis, for instance, accuracy for measuring the exact standardized uptake value (SUV) of a region may be of lesser importance than ensuring that the relative intensity with respect to a reference region (i.e., the SUV ratio, or SUVR) matches that of the acquired image. More importantly for epilepsy patients, these metrics may fail to capture clinically relevant indicators such as the hemispheric asymmetry in metabolism, which is important for downstream tasks such as epilepsy diagnosis and classification.

In this study, we compared state-of-the-art deep learning diffusion and transformer models for the specific task of translating MRI and MRI plus ultralow dose FDG-PET to full-dose FDG-PET in epilepsy patients. Using 2D models, we go beyond earlier works that focused only on T1-weighted (T1w) images, also incorporating other MR contrasts, such as T2 FLAIR and 1\% ultralow-dose PET as training inputs, demonstrating the potential for reduced radiation exposure. Finally, we proposed and evaluated the models on metrics with clinical relevance for epilepsy patients.

\section{Material and Methods}
\subsection{Diffusion Models in Medical Imaging}
DDPMs have captured significant attention in recent years as a class of generative models that generate data by learning to reverse a diffusion process. These models operate in two stages: in the forward process, the original data is gradually corrupted with Gaussian noise over multiple timesteps, eventually becoming pure noise. In the backward process, the model learns to reverse this transformation step-by-step, recovering the original data distribution from the noise. Conditional inputs such as prior image or textual context can be passed into the network for all timesteps This framework has been actively explored in medical imaging due to its ability to capture complex data distributions. For instance, a DDPM with fast training and sampling was applied to T1w-to-T2 image translation using the Brain Tumor Segmentation (BraTS) dataset, with performance evaluated through metrics like PSNR and SSIM \cite{fastddpm}. Additionally, DDPMs have shown promise in CT-to-MRI image translation \cite{ct_mri_ddpm} and for PET denoising \cite{Singh_2024, Gong2024}. However, based on our initial exploration, DDPMs / Latent Diffusion models did not produce satisfactory results with small datasets. 

Closely related to DDPMs, Score-Based Generative Models (SGMs) learn the gradient of the log probability density function of the data of interest, known as the score function. SGMs leverage stochastic differential equations (SDEs) to disturb, model, and sample from the data distribution, using sampling techniques such as Predictor-Corrector, Euler-Maruyama, and Heun’s methods \cite{ct_mri_ddpm, Singh_2024, Gong2024, song2021, karras2022}. 

In the forward process in equation \eqref{eq:1}, the data distribution is disturbed towards Gaussian noise, where change of x at time t \textbf{($dx_{t}$)} consists of a drift term f(x, t), a diffusion term g(t) and a Weiner process term  (white noise) \cite{song2021}. The drift term f can be understood as the deterministic part of the system’s evolution, while the diffusion term g $\cdot$ $dw_{t}$ resembles the stochastic component adding random noise to the system. 
\begin{equation}
dx = f(x,t) + g(t)dw_{t} \tag{1} \label{eq:1}
\end{equation}
In the backward process in equation \eqref{eq:2}, the Gaussian noise is gradually removed from the original data following the backward SDE equation \cite{song2021}. A deep learning network with parameters $\theta$ is used to learn the gradient of the log probability of the data distribution, $\nabla_x \log p_t(x)$, and is referred to as the score  in equation \eqref{eq:3}. Once this gradient is learned, we can obtain our target image from repetitively removing noise through this backward process.
\begin{equation}
dx = \left[f(x,t) - g(t)^2 \nabla_x \log p_{t}(x) dw_{t}\right]dt + g(t)dw_{t} \tag{2} \label{eq:2}
\end{equation}
\begin{equation}
\nabla_x \log p_{t}(x|x_{0}) = s_{\theta}(x, t) \tag{3} \label{eq:3}
\end{equation}
Various score-based models have been used in medical image tasks. SGM with Variance Exploding SDE scheme (VESDE) is applied to CT-to-MRI image translation and achieved better performance than CNN and GAN models for SSIM and PSNR \cite{ct_mri_ddpm}. SGM with Variance Preserving SDE scheme (VPSDE) has been applied to PET reconstruction \cite{Singh_2024} incorporating the use of forward projections of PET images to count measurement into its sampling. Next, we introduce the SGM we used in this work in more detail.

\subsection{Conditional SGM}
VPSDE was used in \cite{Singh_2024}for PET reconstruction task and was selected for evaluation in the current work. VPSDE is a type of SDE with the formulation in equation \eqref{eq:4} in the forward pass, where  is a positive noise scalar with 0 < $\beta_{1}$ < $\beta_{2}$ < .... < $\beta_{T}$ < 1 . At time t, the noise-disturbed x follows a Gaussian distribution with mean of $x(0)e^{-\frac{1}{2} \int_0^t \beta(s) \, ds}$ and variance of $\sigma^{2} = I - Ie^{-\int_0^t \beta(s) \, ds}$ (equation \eqref{eq:5}), and the theoretical gradient of the disturbed data is formulated in equation \eqref{eq:6} where  is the \textbf{true noise} we add in the forward process.
\begin{equation}
    dx = -\frac{1}{2} \beta(t)x\cdot dt + \sqrt{\beta(t)} \cdot dw \tag{4} \label{eq:4}
\end{equation}
\begin{equation}
    p(x_{t}|x_{0})= N(x_{t};x_{0}e^{-\frac{1}{2} \int_0^t \beta(s) \, ds}, I - Ie^{-\int_0^t \beta(s) \, ds}) \tag{5} \label{eq:5}
\end{equation}
\begin{equation}
    \nabla_x \log{p_{t}(x)} = -\frac{x_{t}-x_{0}}{\sigma_{t}^2}=-\frac{x_{t}-(x_{t}-\epsilon\sigma_{t})}{\sigma_{t}^2} = -\frac{\epsilon}{\sigma_{t}} \tag{6} \label{eq:6}
\end{equation}
Our SGM generates PET sample conditioned on the MRI/PET low-dose inputs, so that the learning of the distribution is conditioned on \textbf{y }and we can rewrite equations \eqref{eq:2} and \eqref{eq:3} to equations \eqref{eq:7} and \eqref{eq:8} with the additional input \textbf{y}. The conditional neural network  aims to learn the true gradient $-\frac{\epsilon}{\sigma_{t}}$  in the forward process through minimizing the loss function with conditional inputs (scaled by weighting $\lambda_{t}$) shown by equation \eqref{eq:9}. The loss is averaged over all time points and averaged over all training data x.

\begin{equation}
dx = \left[f(x,t) - g(t)^2 \nabla_x \log p_{t}(x) dw_{t}\right]dt + g(t)dw_{t} \tag{7} \label{eq:7}
\end{equation}
\begin{equation}
\nabla_x \log p_{t}(x|x_{0}, y) = s_{\theta}(x, y, t) \tag{8} \label{eq:8}
\end{equation}
\begin{equation}
    \mathrm{Loss = E_{t}{\lambda_t E_{x_{0}}E_{x_{t}|x_{0},y}[\left\lVert s_\theta(x, y, t) - \nabla_x \log p_{0t}(x_t \mid x_0, y) \right\rVert^2_2
]}} \tag{9} \label{eq:9}
\end{equation}
\begin{equation}
    \mathrm{Loss = E_{t}{\lambda_t E_{x_{0}}E_{x_{t}|x_{0},y}[\left\lVert s_\theta(x, y, t) + \frac{\epsilon}{\sigma_{t}} \right\rVert^2_2
]}} \tag{10} \label{eq:10}
\end{equation}
For the backward sampling process, our sampler follows a predictor and corrector style from \cite{song2021} with the following algorithm. Starting with  from Gaussian noise, the predictor uses the score from the network combining with equation \eqref{eq:7} to iteratively update x. After each predictor step there is a fixed number of corrector steps that uses a normalized and scaled score to further improve the current x. The overall workflow of the reverse process is shown in Figure \ref{fig:1}a. At each timestep, the noisy prediction, the timestep information, and the conditional inputs are passed to the neural network, which outputs the gradient of the current prediction used for updates and corrections.

\begin{algorithm}
\caption{VPSDE Sampling}
\label{alg:noise_to_image}
\begin{algorithmic}[1]
\State $x_N \sim \mathcal{N}(0, I)$ \Comment{Starts with random noise}
\State $y \gets \text{MRI/PET 1\% (condition)}$
\For{$T = N-1 \text{ to } 1$} \Comment{Predictor step}
    \State $\text{score} \gets s_\theta(x_{T+1}, y, t_{T+1})$
    \State $\Delta t \gets t_{T+1} - t_T$
    \State $z \sim \mathcal{N}(0, I)$
    \State $x_T \gets x_{T+1} + \left[ f(x_{T+1}, t_{T+1}) - g(t_{T+1})^2 \cdot \text{score} \right] \Delta t$
    \State \hspace{1cm} $+ \; g(t_{T+1}) \sqrt{|\Delta t|} \cdot z$
    \For{$j = 1 \text{ to } M$} \Comment{Corrector step}
        \State $z \sim \mathcal{N}(0, I)$
        \State $g \gets s_\theta(x_T^{(j-1)}, y, t_T)$
        \State $\epsilon \gets 2\alpha \left( \frac{\|z\|_2}{\|g\|_2} \right)^2$ \Comment{Normalize the score $g$}
        \State $x_T^{(j)} \gets x_T^{(j-1)} + \epsilon \cdot g + \sqrt{2\epsilon} \cdot z$
    \EndFor
\EndFor
\State \Return $x_0$
\end{algorithmic}
\end{algorithm}

Another proposed architecture using SGM comes from Karras et al. \cite{karras2022}, which we abbreviate as SGM-KD), which envisions a denoiser D reconciling different SDE schemes with the following formulation:
\begin{equation}
    D_{\theta}(x, \sigma)=c_{skip}(\sigma)x + c_{out}(\sigma)F_{\theta}(c_{in}(\sigma)x; c_{noise}(\sigma)) \tag{11} \label{eq:11}
\end{equation}
Here $\sigma$ is the noise level at a specific timestep. $F_{\theta}$ is the neural network to be trained on noise-disturbed data, $c_{skip}(\sigma)$  is a $\sigma$-dependent scalar modulating the skip connection,  $c_{in}(\sigma)$ and  $c_{out}(\sigma)$ modulates the input and output magnitude and $c_{noise}(\sigma)$ is a scalar served as a conditional input into the network. In our task, y is the MRI/PET low-dose condition, and x is the full-dose PET data, we can add our conditions into the equation:
\begin{equation}
    D_{\theta}(x, \sigma, y)=c_{skip}(\sigma)x + c_{out}(\sigma)F_{\theta}(c_{in}(\sigma)x; c_{noise}(\sigma), y) \tag{12} \label{eq:12}
\end{equation}
The denoiser aims to minimize the difference between the denoised output with the target x, which can be regarded as predicting and removing the noise added at different steps.
\begin{equation}
    \mathrm{Loss = E_{\sigma, PET, noise, MRI}[\lambda(\sigma)\left\lVert D(PET_{clean} + noise; \sigma, MRI) - PET_{clean}\right\rVert]} \tag{13} \label{eq:13}
\end{equation}

During the sampling process, we used the paper’s stochastic sampler, which averages the gradient from the current timestep and a future timestep for improved quality:

\begin{algorithm}
\caption{Stochastic Sampler with $\sigma(t) = t$, $s(t) = 1$ and conditional denoiser $D_{\theta}$}
\begin{algorithmic}[1]
\State Sample $x_0 \sim \mathcal{N}(0, t_0^2 \mathbf{I})$
\State $y \gets \text{MRI/PET 1\% (condition)}$
\For{$i \gets 0$ to $N-1$}
    \State Sample $\epsilon_i \sim \mathcal{N}(0, S_{\text{noise}}^2 \mathbf{I})$
    \State $t_i \gets t_i + \gamma_i t_i$
    \State $\hat{x}_i \gets x_i + \sqrt{t_i^2 - t_i'^2} \epsilon_i$
    \State $d_i \gets (\hat{x}_i - D_\theta(\hat{x}_i, t_i, y)) / t_i$
    \State $x_{i+1} \gets \hat{x}_i + (t_{i+1} - t_i) d_i$
    \If{$t_{i+1} \neq 0$}
        \State $d_i' \gets (x_{i+1} - D_\theta(x_{i+1}, t_{i+1}, y) / t_{i+1}$
        \State $x_{i+1} \gets \hat{x}_i + (t_{i+1} - t_i) \left(\frac{1}{2} d_i + \frac{1}{2} d_i'\right)$
    \EndIf
\EndFor
\State \Return $x_N$
\end{algorithmic}
\end{algorithm}

\subsection{Transformer-Unet (TransUnet)}
\begin{figure}
    \centering
    \includegraphics[width=1\linewidth]{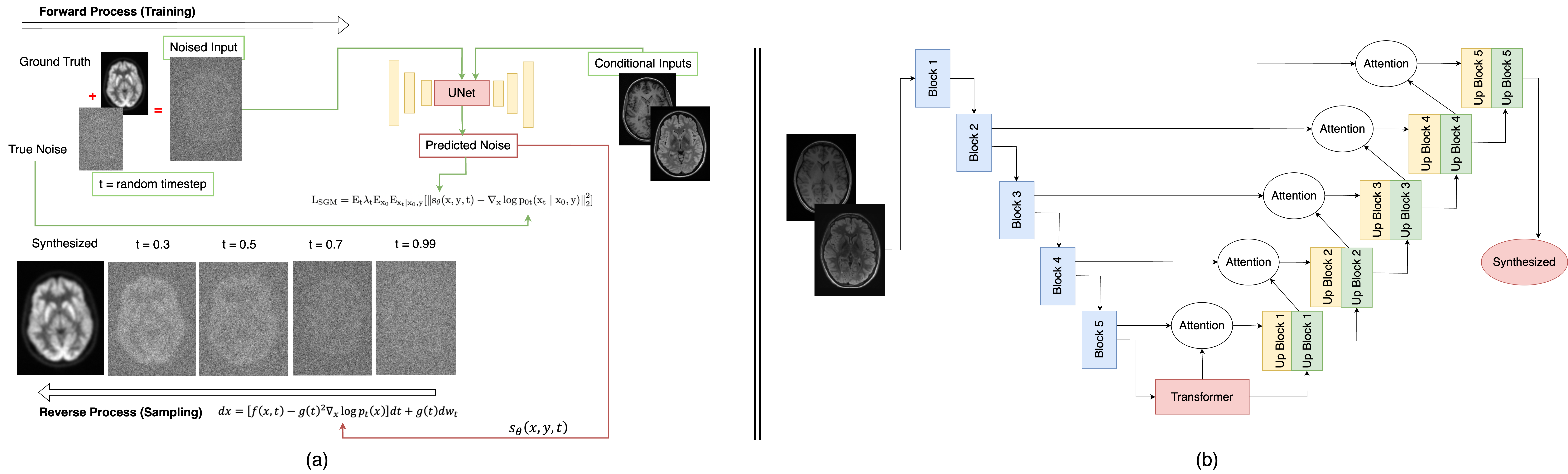}
    \caption{Visualization of SGM (a) and TransUnet (b) architectures. In SGM, training occurs on the forward process, where random timestep + noisy inputs + conditional inputs are used to minimize the loss between true and predicted noise (green boxes/arrows). Sampling starts from pure random noise. At every timestep t, the timestep + noisy predictions + conditional inputs are passed to the model. The model outputs a prediction on the gradient of the data distribution at current timestep, which is used to update the prediction (red arrow). In TransUnet, the blue blocks are the downsampling blocks. The yellow and green blocks are different inputs at each layer and are concatenated together for the subsequent upsampling procedure. Channel and spatial attention are also added to each layer.}
    \label{fig:1}
\end{figure}

TransUnet, developed in our previous work \cite{transunet}, takes in multi-contrast MRI inputs to predict full-dose FDG-PET and demonstrated strong performance based on metabolic metrics such as SUVR. As shown in Figure \ref{fig:1}b, the multi-contrast inputs are concatenated at the start of the network. There are 5 blocks with batch-normalization and convolution layers with stride of 2 so that the 256x256 inputs are downsampled to 8x8. Next, a transformer block is applied on the flattened inputs at this innermost layer with multi-head self-attention to capture global information. During the upsampling blocks, spatial attention, channel attention and upsampling are applied to the inputs and concatenated with the corresponding inputs from the downsampling blocks. Finally, an output convolution layer is applied to obtain the final synthetic PET slice. 

\subsection{Patient Population and Imaging Methods}
After receiving a waiver of consent from our IRB, we identified consecutive adult patients who underwent simultaneous FDG 3.0T brain PET/MRI (GE Signa, Waukesha, WI) examinations for clinical purposes between August 2022 and October 2024 with available list-mode data. The primary clinical indication for these studies was seizure evaluation, though cases with other indications (e.g., dementia, tumor) were also present. However, only patients undergoing evaluation for epilepsy were included in the test sets, reflecting our specific focus on epilepsy in this study. All available cases, including non-epilepsy cases, were included in the training set to maximize dataset size. Dataset splits and input details are shown in Figure \ref{fig:2}, where we identified 52 cases (9 non-epilepsy and 43 epilepsy cases).  Ten and two epilepsy patients were randomly selected as test and validation cases, respectively.  The remainder of the patients were used for model training.

All patients received a standard full-dose radiotracer, approximately 3.0 MBq/kg.  PET counts were acquired for 10 min following a 40 min uptake time.  For both the 100\% and ultralow-dose PET images, the reconstruction method was ordered subset expectation maximization using 28 subsets and 4/1 iterations. A 4 mm Gaussian smoothing filter was applied to all images.  Ultralow-dose PET images were created using list-mode random count under-sampling. Specifically, for all coincidence events recorded in the list-mode data, we selected 1 event every 100 events for 1\% PET simulation and passed the selected events into the 3D PET reconstruction \cite{chen2020ultra}.  The dimensions of each reconstructed PET was 256 x 256 x 89 with nominal resolution of 1.2 x 1.2 x 2.78 mm. High-resolution 3D T1w and T2 FLAIR images were acquired using our clinical protocol with the following parameters: T1: TR 6-8.5ms, TI = 400-600ms, TE = 2.8ms, flip angle = 12°, slice thickness = 1.5mm; T2 FLAIR: TR 5602-6002ms, TI = 1567-1765ms, TE = 127ms, flip angle = 90°, slice thickness = 2mm. 

\begin{table}[h]
\begin{tabular}{@{}lll@{}}
\toprule
 & Training/Validation (n=42)  & Testing (n=10) \\
\midrule
Age    & 42.2 $\pm$ 15.1   & 34.6 $\pm$ 11.1\\
Sex    & 18F, 24M & 6F, 4M  \\
Radiotracer Dose Level    & 2.85 $\pm$ 0.65 MBq/kg  & 3.00 $\pm$ 0.73 MBq/kg \\
\botrule
\end{tabular}
\caption{\centering Table for training and testing data cases}
\label{tab:demographics}
\end{table}

\subsection{Implementation Detail}
We evaluated a TransUnet and the 2 SGMs introduced earlier due to their stable synthetic quality and efficient sampling. Specifically, we selected SGMs with variance-preserving SDE with predictor-corrector sampler (referred to as SGM-VP) used in a prior PET reconstruction work \cite{Singh_2024} and SGM with 2nd order sampler as proposed by Karras et al. (referred to as SGM-KD) \cite{karras2022}. For the SGMs, the data is normalized to a range of [0, 1] for SGM-VP and [-1, 1] for SGM-KD. For TransUnet, we follow the architecture described in \cite{transunet} and normalize by dividing all voxel values by the mean value of the volume. All MRI volumes are co-registered to PET native space to avoid information loss.

\begin{figure}[H]
    \centering
    \includegraphics[width=1.0\linewidth]{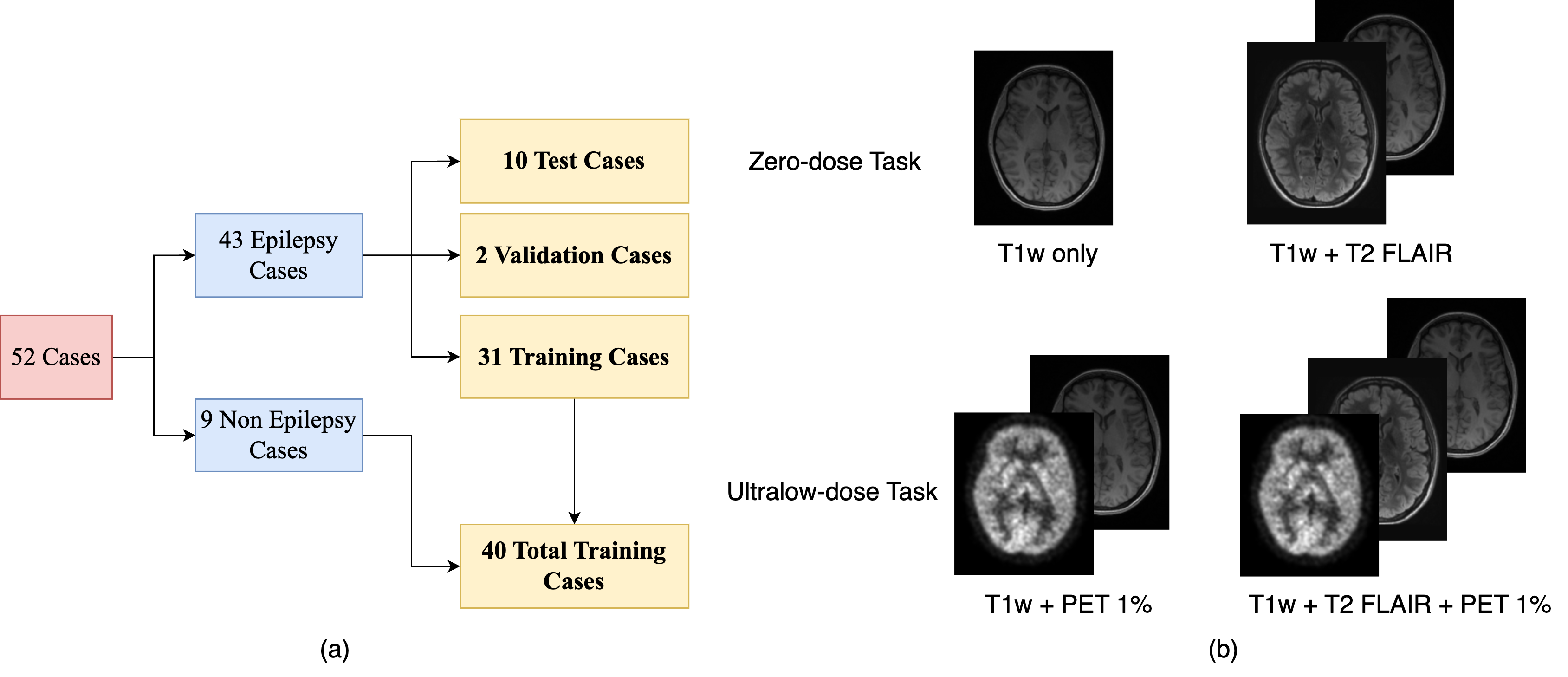}
    \caption{(a) Visualization of data split between training, validation, and test sets.  (b) 4 different input combinations explored in this study. }
    \label{fig:2}
\end{figure}

We investigated the effect of multi-sequence MRI and ultralow-dose PET using 4 different input combinations for each model: T1w only (no PET data), T1w + T2-FLAIR (no PET data), and 1\% PET data combined with either T1w or T1w + T2-FLAIR (Figure \ref{fig:2}b). For all models, we used 3 MRI slices to predict the middle PET slice.  Larger numbers of surrounding slices were explored but showed limited improvement. Training was done on a 32GB GPU (Quadro GV100, Tesla V100-PCIE) with the largest batch size allowed by the memory (8-16 per batch for SGMs, 64 per batch for TransUnet). 

\subsection{Evaluation Metrics}
For evaluation, brain segmentation was performed using Freesurfer’s Destrieux parcellation \cite{DESTRIEUX2009S151}, and related small regions belonging to the same overarching larger region were grouped together (e.g., all sub-regions within frontal cortex were grouped into frontal cortex). We performed measurements in 10 region-of-interests (ROI): frontal cortex (\textbf{FC}), temporal cortex (\textbf{TC}), parietal cortex (\textbf{PC}), occipital cortex (\textbf{OC}), insular cortex (\textbf{IC}) cerebral white matter (\textbf{CWM}), deep gray matter including caudate, putamen, and globus pallidus (\textbf{DGM}), combined hippocampus and amygdala (\textbf{HipAmy}), cerebrospinal fluid (\textbf{CSF}), and cerebellum. Standardized Uptake Value Ratios (SUVR) were then calculated using the cerebellum as the reference region. The mean absolute difference in SUVR ($\Delta$SUVR Mean) between the acquired PET and synthetic PET is computed across the whole brain volume across all test subjects, and the standard deviation of the voxel-wise difference ($\Delta$SUVR STD) across brain volume is also computed individually and averaged across all test subjects. This evaluates the accuracy and consistency of model performance on a voxel level.

Given the diagnostic importance of hemispheric metabolic differences in epilepsy \cite{LU201353}, we calculated a clinically relevant metric termed the Congruence Index (CI)to assess the agreement of SUVR asymmetry between the synthesized and the acquired PET images. We first construct an Asymmetry Index (AI) \cite{Soma2012, Didelot1732, flowgan} for each ROI using equation \eqref{eq:14}, where the difference in SUVR from the left and right hemisphere is divided by the sum. As such, a positive AI indicates right-sided hypometabolism while a negative AI indicates left-sided hypometabolism \cite{flowgan}. 
\begin{equation}
    \mathrm{AI_{ROI}(PET) = \frac{SUVR_{left ROI}(PET) - SUVR_{right ROI}(PET)}{SUVR_{left ROI}(PET) + SUVR_{right ROI}(PET)}} \tag{14} \label{eq:14}
\end{equation}
Next, we constructed a Congruence Index using formulation in below to obtain the proportion of ROIs matching the asymmetry present in the acquired PET. We identified 8 areas with laterality: FC, TC, OC, IC, PC, CWM, DGM, and HipAmy. We averaged the index across the 10 test subjects. A higher CI indicates better representation of the left-right asymmetries found in the reference full-dose image. The formulation for CI calculation is summarized in \eqref{eq:15}:
\begin{equation}
    \mathrm{CI_{synth} = \frac{1}{N_{subjects}N_{ROIs}}\sum_{subjects, ROIs} [[ AI_{ROI}(PET_{synth}) \times AI_{ROI}(PET_{acquired}) > 0]]} \tag{15} \label{eq:15}
\end{equation}
Considering the different ROI sizes and the observation that small asymmetries may not be captured by human readers, we further included the absolute errors and penalized more on the mismatch of larger ROIs using a scale $\mathrm{\frac{Area_{ROI}}{Area_{max}}}$ ($\mathrm{Area_{max}}$ is usually the cerebral white matter). Thus, we included a Congruence Mean Absolute Error (CMAE) averaged across 10 subjects where a smaller value indicates a more accurate synthesis in \eqref{eq:16}:
\begin{equation}
    \mathrm{CMAE = \frac{1}{N_{subjects}}\sum_{subjects, ROIs}|AI_{ROI}(PET_{acquired})-AI_{ROI}(PET_{synth})| \times \frac{Area_{ROI}}{Area_{max}}} \tag{16} \label{eq:16}
\end{equation}

To analyze slice inconsistencies that can occur using 2D models for 3D image generation, we evaluated randomly selected slices reconstructed in the coronal orientation. From this, we computed the line profile of signal intensity along the cranial-caudad direction for the synthesized and acquired PET studies.

\subsection{Statistical Analysis}
Since SUVR is both subject- and ROI-based, we employed a mixed-effects model to account for both between-group and within-group variance. Using the variances estimated by this model, we calculated the Intraclass Correlation Coefficient (ICC) to assess the reliability of SUVR values from the synthetic PET images compared to the ground truth. According to \cite{icc}, ICC values between 0.5 and 0.75 indicate moderate reliability, values between 0.75 and 0.9 indicate good reliability, and higher ICC values reflect greater reliability.

\section{Results}
\subsection{Visual Quality of Synthetic Full-dose FDG-PET}
\begin{figure}
    \centering
    \includegraphics[width=1.0\linewidth]{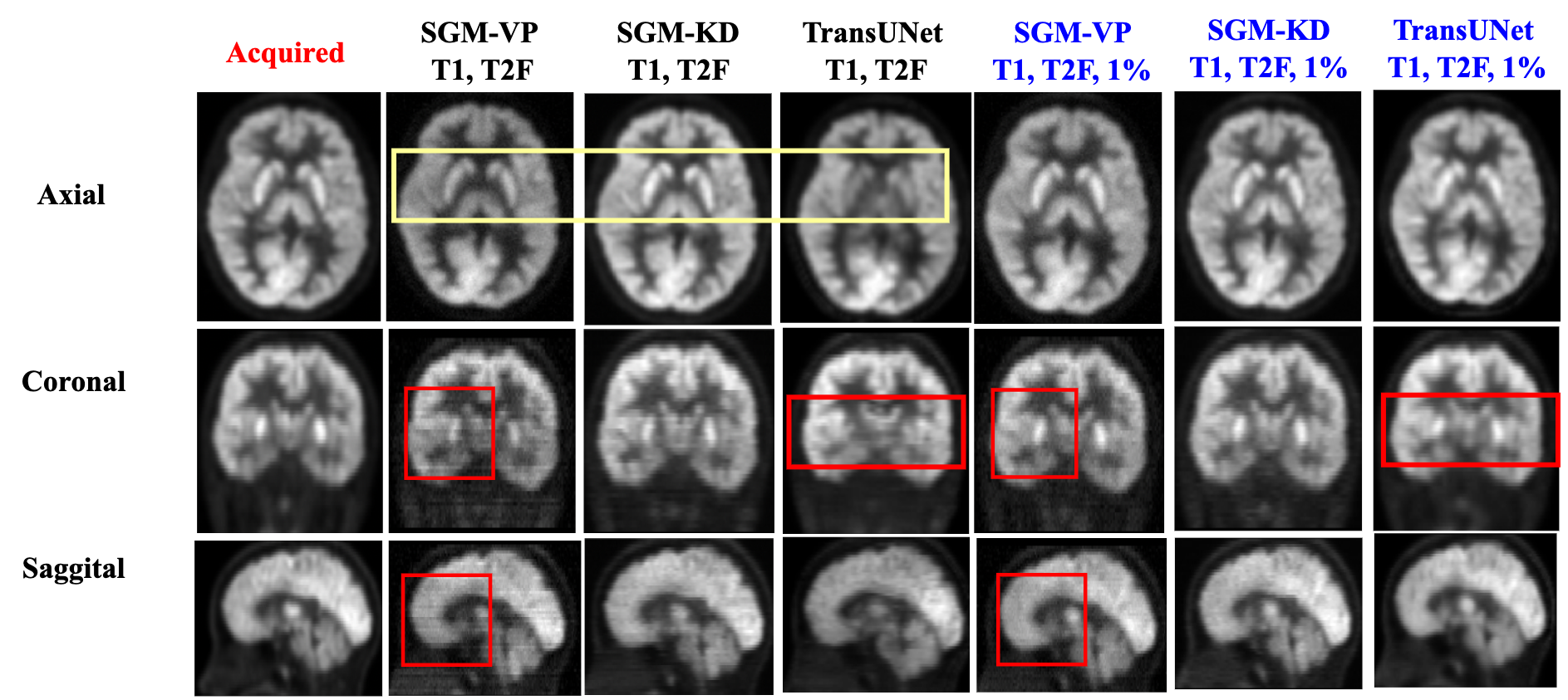}
    \caption{Slice visualization for a test subject in all 3 views for different models and different inputs. SGM-VP and SGM-KD are score-based diffusion models with variations in the noise scheduling. The yellow box compares the anatomical details generated by 3 models in zero-dose task, from which we observe less accurate detail from TransUnet, particularly in the region of the basal ganglia. The red box indicates the effect of the 1\% PET input in improving the slice consistency in coronal and sagittal view and anatomical details (moving from black columns to blue columns). This improvement is the most significant for TransUnet, emphasizing the importance of ultralow-dose inputs.}
    \label{fig:3}
\end{figure}

\begin{figure}
    \centering
    \includegraphics[width=0.7\linewidth]{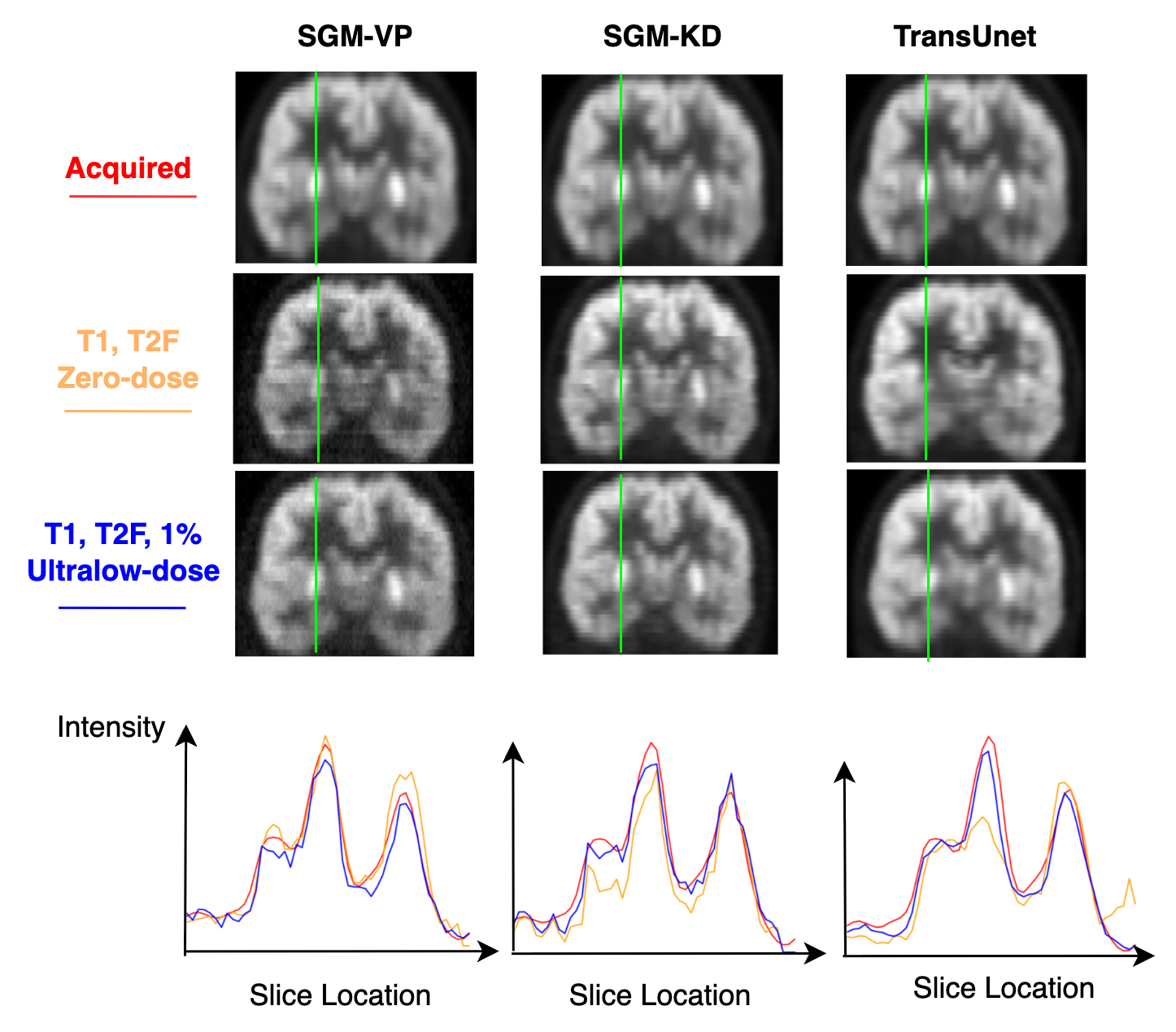}
    \caption{Interslice consistency evaluation between zero-dose and ultralow-dose tasks using the line profile from 1 coronal slice in each PET (intensity values taken along the green lines). On the bottom graphs, the red curve represents the acquired PET, the orange curve represents the zero-dose task, and the blue curve represents the ultralow-dose task. The blue curve is closer to the red curve, and has less high-frequency components, indicating improved slice intensity consistency for the ultralow-dose task.}
    \label{fig:4}
\end{figure}

For the zero-dose task using only MRI sequences as inputs, the 2 SGM models outperformed the TransUnet, as shown by the more accurate synthesis in the region highlighted by the yellow boxes in Figure \ref{fig:3}. The TransUnet (CNN style network) tended to generate oversmoothed images with fewer details, consistent with the observation made in \cite{ct_mri_ddpm}, whereas score-based models emphasize the edges and the contrast between regions with low and high metabolism. For the ultralow-dose task that included 1\% dose PET images as input, the two SGM models exhibited reduced slice-to-slice inconsistencies in coronal and sagittal views, improving the quality of coronal and sagittal views (Figure \ref{fig:3}).  This is also seen in Figure \ref{fig:4}, which provides a closer examination of the slice inconsistency improvement moving from zero-dose to ultralow-dose task. The ultralow-dose line profiles exhibit less jittering compared with the zero-dose profiles, reflecting smoother intensity transitions between slices that more closely resemble the acquired PET. Notably, the TransUnet shows the least slice inconsistency in the zero-dose task, but requires 1\% PET input to generate sufficient anatomical detail. This limitation may stem from its reduced complexity, which could hinder performance with limited training data.

\begin{figure}[H]
    \centering
    \includegraphics[width=0.8\linewidth]{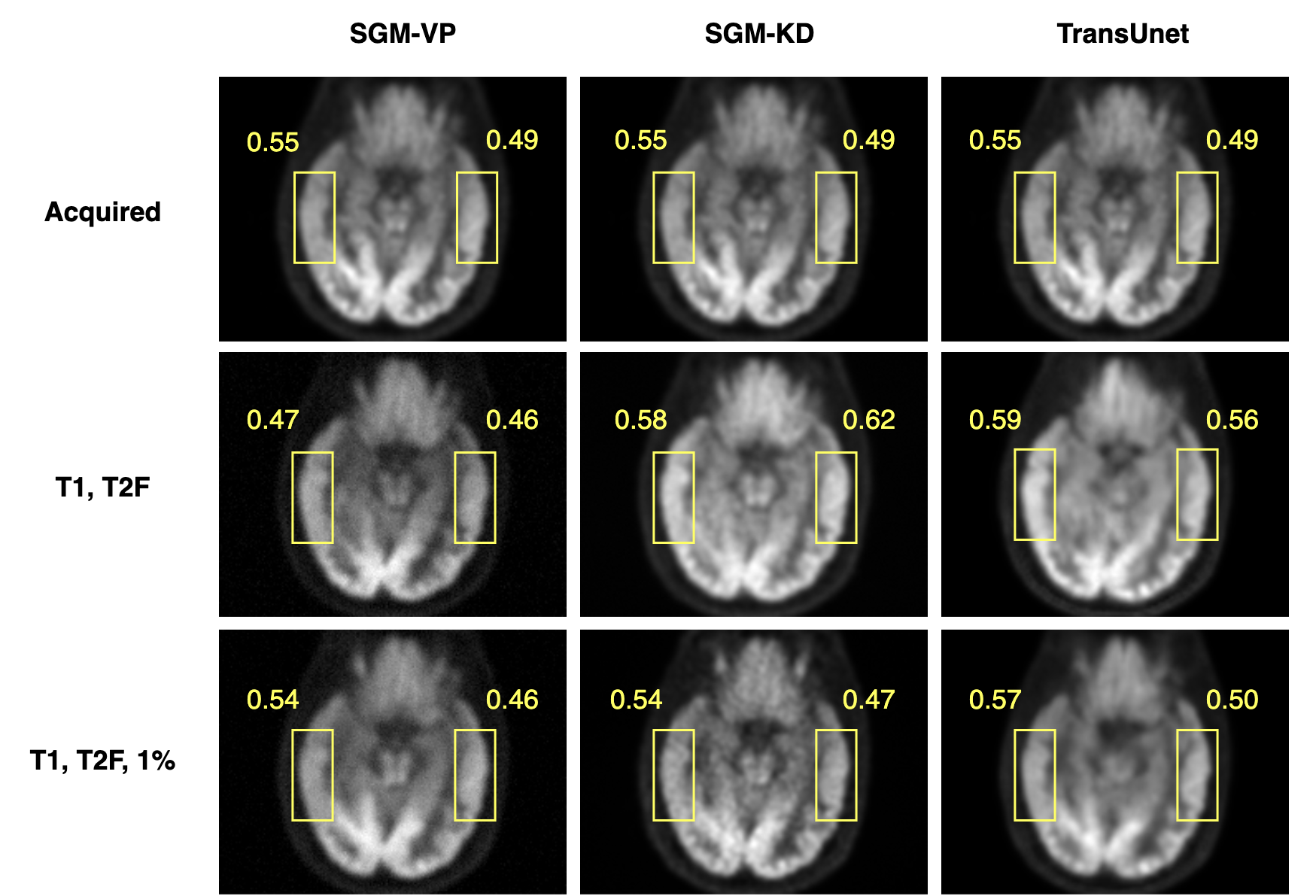}
    \caption{Axial slice for one case with asymmetrically lower metabolism in the left temporal lobe based on the clinical report. We computed the mean SUVR in the left and right yellow boxes with values shown on top. We observe better agreement with the acquired PET for ultralow-dose task compared to zero-dose task. }
    \label{fig:5}
\end{figure}

Lastly, we examined the quality of the synthetic PET for a test case with left-sided temporal lobe hypometabolism based on the clinical reports. As shown in Figure \ref{fig:5}, the left-right asymmetries are much easier to identify on the ultralow-dose synthesis as compared to the zero-dose synthesis.  This is true for all three models. 

\subsection{Quantitative Performances}
Next, we evaluated the quantitative metabolic aspects of the synthesized images for zero-dose task (Table \ref{tab:zerodose1}). The SGM-VP model demonstrated the best performance for the asymmetry metrics CI and CMAE, indicating that SGM-VP is more accurate for larger proportion of the test subjects and ROIs. The ICC for all models is around 0.75 or greater, suggesting that they all have good reliability. While ICCs are all relatively comparable, the SGM-KD model using T1w and T2-FLAIR as inputs had the highest ICC (0.84).

\begin{table}[h!]
\begin{tabular*}{\textwidth}{@{\extracolsep{\fill}}ccccccccc}
\toprule
\begin{tabular}[c]{@{}c@{}}Model\\ Type\end{tabular} & Inputs & \begin{tabular}[c]{@{}c@{}}SUVR\\ ICC\end{tabular} & \begin{tabular}[c]{@{}c@{}}SUVR\\ ICC CI\end{tabular} & \begin{tabular}[c]{@{}c@{}}CMAE\\(x$10^3$)\end{tabular}        & \begin{tabular}[c]{@{}c@{}}CMAE\\ CI (x$10^3$)\end{tabular} & \begin{tabular}[c]{@{}c@{}}Congruence\\ Index \end{tabular} & \begin{tabular}[c]{@{}c@{}}Congruence \\ Index CI \end{tabular} \\ \hline 
SGM-VP & T1w & 0.81 & 0.65, 0.91 & \textbf{0.61} & 0.23, 0.99 & \textbf{0.85} & 0.73, 0.97 \\ \hline
SGM-VP & \begin{tabular}[c]{@{}c@{}}T1w,\\   T2F\end{tabular} & 0.80 & 0.63, 0.90 & 1.11 & 0.67, 1.56 & 0.76 & 0.70, 0.83 \\ \hline
SGM-KD & T1w & 0.82 & 0.67, 0.91 & 1.10 & 0.71, 1.49 & 0.71 & 0.60, 0.82 \\ \hline
SGM-KD & \begin{tabular}[c]{@{}c@{}}T1w, \\ T2F\end{tabular}  & \textbf{0.84} & 0.70, 0.92& 0.73 & 0.30, 1.15 & 0.76 & 0.66, 0.87 \\ \hline
TransUnet & T1w & 0.80 & 0.63, 0.90 & 1.01 & 0.48, 1.54 & 0.74 & 0.62, 0.86 \\ \hline
TransUnet & \begin{tabular}[c]{@{}c@{}}T1w,\\   T2F\end{tabular} & 0.74 & 0.55, 0.87 & 1.07 & 0.45, 1.70 & 0.61 & 0.51, 0.71\\ \hline
\end{tabular*}
\caption{Performance metrics for zero-dose task. The best performance for each metric is shown in bold. The 95\% confidence interval for Congruence Index and Congruence Mean Absolute Error was calculated using t-distribution. CMAE is scaled by 1000 for better visualization.}
\label{tab:zerodose1}
\end{table}

\begin{figure}[H]
    \centering
    \includegraphics[width=1.0\linewidth]{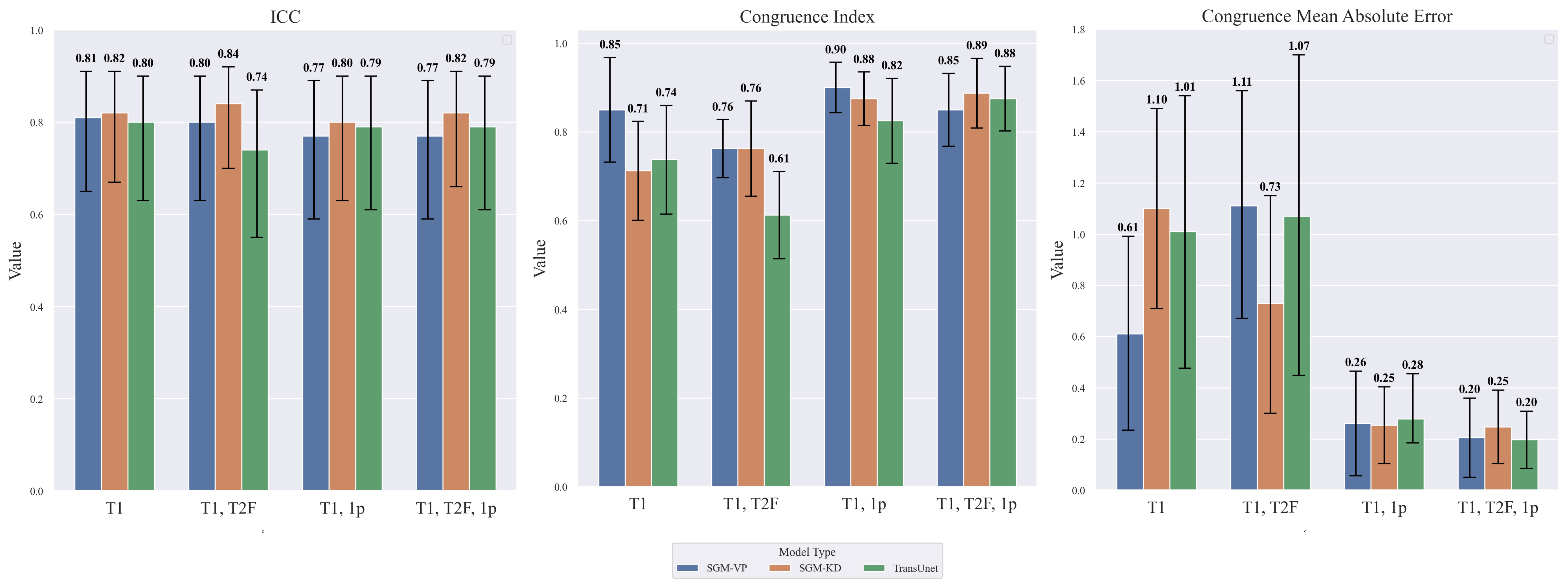}
    \caption{Visualization of ICC, CI, CMAE for the different models and image input conditions.  No significant change in ICC was observed between the zero-dose task and ultralow-dose tasks. CI and especially CMAE demonstrate more significant improvement for models that include 1\% dose PET as an input.  Please note that higher CI and lower CMAE indicates better performance.}
    \label{fig:6}
\end{figure}

When PET 1\% data was included as input, there was improved performance for the asymmetry indices across all model types (Table \ref{tab:lowdose1}). Specifically, the best performing CI increases to a high of 0.9 while the best CMAE is 0.20. TransUnet experiences the largest improvement compared to SGMs and is now capable of capturing smaller anatomical details (see Figure \ref{fig:3}). ICC values were largely unchanged, still exceeding the 0.75 threshold indicative of good reliability, suggesting that ICC is less dependent on the presence of ultralow-dose inputs. Figure \ref{fig:6} presents these findings in a summary chart that allows clear visualization of the improvement of the CI and CMAE metrics for all models when the 1\% PET image is included as an input.   

\begin{table}[h!]
\begin{tabular*}{\textwidth}{@{\extracolsep{\fill}}ccccccccc}
\toprule
\begin{tabular}[c]{@{}c@{}}Model\\ Type\end{tabular} & Inputs & \begin{tabular}[c]{@{}c@{}}SUVR\\ ICC\end{tabular} & \begin{tabular}[c]{@{}c@{}}SUVR\\ ICC CI\end{tabular} & \begin{tabular}[c]{@{}c@{}}CMAE\\(x$10^3$)\end{tabular}          & \begin{tabular}[c]{@{}c@{}}CMAE\\ CI (x$10^3$)\end{tabular} & \begin{tabular}[c]{@{}c@{}}Congruence\\ Index \end{tabular} & \begin{tabular}[c]{@{}c@{}}Congruence \\ Index CI \end{tabular} \\ \hline 
SGM-VP & T1w, 1\% & 0.77 & 0.59, 0.89 & 0.26 & 0.06, 0.47 & \textbf{0.90} & 0.84, 0.96 \\ \hline
SGM-VP & \begin{tabular}[c]{@{}c@{}}T1w,\\T2F, 1\%\end{tabular} & 0.77 & 0.59, 0.89 & 0.21 & 0.05, 0.36 & 0.85 & 0.77, 0.93 \\ \hline
SGM-KD & T1w, 1\% & 0.80 & 0.63, 0.90 & 0.25 & 0.10, 0.40 & 0.88 & 0.82, 0.94 \\ \hline
SGM-KD & \begin{tabular}[c]{@{}c@{}}T1w, \\T2F, 1\%\end{tabular}  & \textbf{0.82} & 0.66, 0.91 & 0.25 & 0.10, 0.40 & 0.88 & 0.82, 0.94 \\ \hline
TransUnet & T1w, 1\% & 0.79 & 0.61, 0.90 & 0.28 & 0.18, 0.45 & 0.83 & 0.73, 0.92 \\ \hline
TransUnet & \begin{tabular}[c]{@{}c@{}}T1w,\\T2F, 1\%\end{tabular} & 0.79 & 0.61, 0.90 & 0.20 & 0.09, 0.31 & 0.88 & 0.80, 0.95\\ \hline
\end{tabular*}
\caption{Performance metrics for ultralow-dose task. The best performance for each metric is in bold. The 95\% confidence interval for Congruence Index and Congruence Mean Absolute Error was calculated using t-distribution. CMAE is scaled by 1000 for better visualization.}
\label{tab:lowdose1}
\end{table}

\begin{figure}[H]
    \centering
    \includegraphics[width=0.8\linewidth]{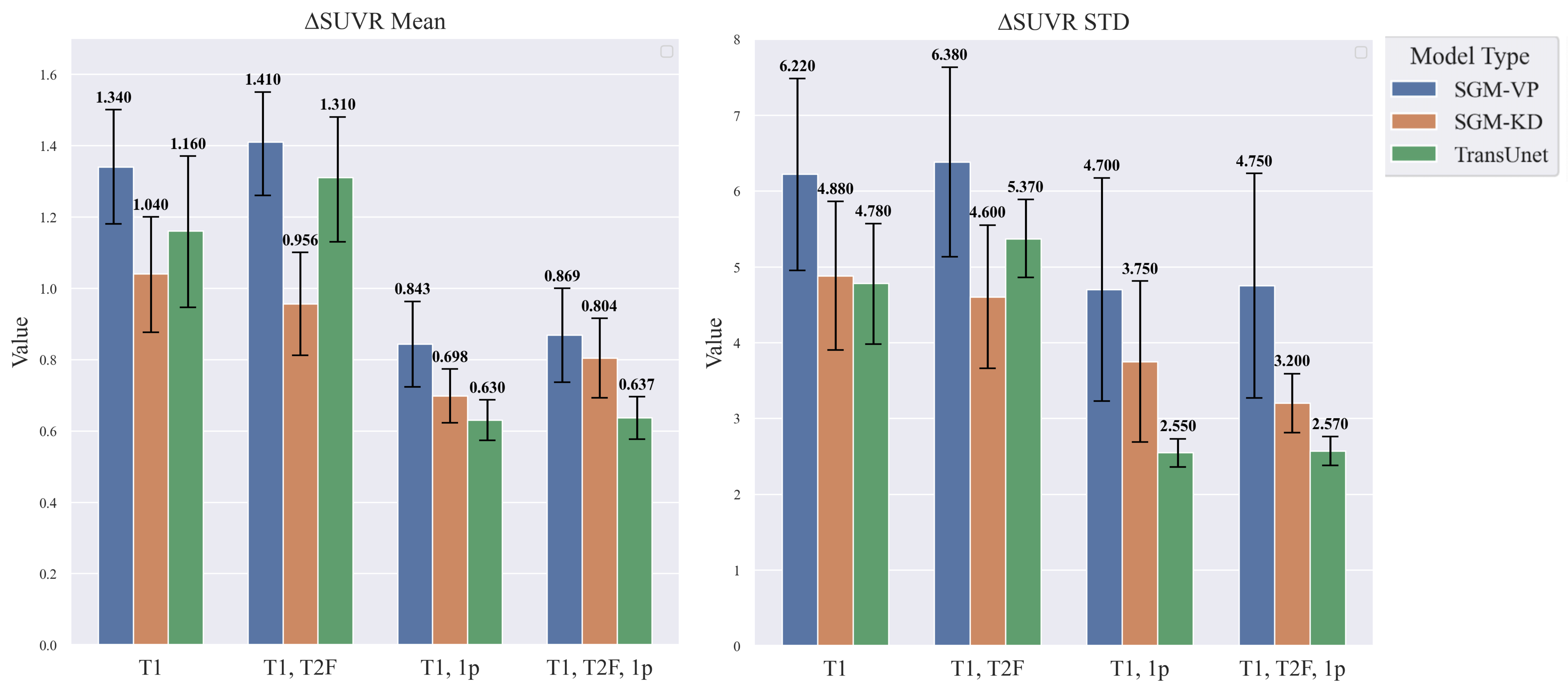}
    \caption{Visualization of for $\Delta$SUVR Mean and STD for different models and image input conditions. There is decreased $\Delta$SUVR mean and std from zero-dose to ultralow-dose, indicating higher voxel-wise accuracies and consistencies.}
    \label{fig:7}
\end{figure}

Lastly, the performances on mean voxel-wise $\Delta$SUVR and mean $\Delta$SUVR standard deviation are shown in Table \ref{tab:suvr} and Figure \ref{fig:7}, where the top half represents the zero-dose task and the lower half the ultralow-dose task. For the zero-dose task, SGM-KD showed both the lowest $\Delta$SUVR Mean and STD, meaning that although SGM-KD is not the best for asymmetry measures, it has a more balanced performance across the volume. For the ultralow-dose task, TransUnet has the lowest $\Delta$SUVR Mean and STD. We notice that although SGM-VP performs better for asymmetry measures, SGM-KD outperforms in the $\Delta$SUVR metrics. A possible explanation is the limited dataset size, which may introduce differences that would not otherwise occur. Additionally, the architecture of SGM-KD might be more effective at synthesizing pixels with average intensities, rather than those representing hypo- or hyper-metabolized regions at the distribution’s extremes. However, these performance differences may change when more data becomes available.

\begin{table}[]
\begin{tabular}{cccccc}
\hline
\multicolumn{1}{c}{\begin{tabular}[c]{@{}c@{}}Model\\ Type\end{tabular}} & \multicolumn{1}{c}{Inputs} & \multicolumn{1}{c}{\begin{tabular}[c]{@{}c@{}}$\Delta$SUVR\\ Mean (x100)\end{tabular}} & \multicolumn{1}{c}{\begin{tabular}[c]{@{}c@{}}$\Delta$SUVR\\ Mean CI (x100)\end{tabular}} & \multicolumn{1}{c}{\begin{tabular}[c]{@{}c@{}}$\Delta$SUVR \\ STD (x100)\end{tabular}} & \multicolumn{1}{c}{\begin{tabular}[c]{@{}c@{}}$\Delta$SUVR\\ STD CI (x100)\end{tabular}} \\ \hline
SGM-VP & T1w & 1.34 & 1.18, 1.50 & 6.22 &4.95, 7.48 \\ \hline
SGM-VP & T1w, T2F & 1.41 & 1.26,   1.55 & 6.38  & 5.13, 7.63 \\ \hline
SGM-KD & T1w & 1.04 & 0.88,   1.20 & 4.88 & 3.90, 5.86 \\ \hline
SGM-KD & T1w, T2F  & \textbf{0.96} & 0.81, 1.10 &\textbf{4.60}  & 3.66,   5.55 \\ \hline
TransUnet  & T1w & 1.16 & 0.95, 1.37 & 4.78 & 3.98,   5.57 \\ \hline
TransUnet & T1w, T2F & 1.31 & 1.13,   1.48 & 5.37  & 4.86,   5.89 \\ \hline
\noalign{\hrule height 1.5pt}
SGM-VP & T1w, 1\%  & 0.84 & 0.72,   0.96  & 4.70 & 3.23,   6.17 \\ \hline
SGM-VP & \begin{tabular}[c]{@{}l@{}}T1w, \\ T2F, 1\%\end{tabular} & 0.87 & 0.74, 1.00 & 4.75 & 3.27,   6.23  \\ \hline
SGM-KD & T1w, 1\% & 0.70 & 0.62,   0.77 & 3.75 & 2.69,   4.81\\ \hline
SGM-KD & \begin{tabular}[c]{@{}l@{}}T1w, \\ T2F, 1\%\end{tabular} & 0.80 & 0.69, 0.92 & 3.20 & 2.81,   3.59 \\ \hline
TransUnet & T1w, 1\% & \textbf{0.63} & 0.57,   0.69 & \textbf{2.55} & 2.36,   2.73 \\ \hline
TransUnet & \begin{tabular}[c]{@{}l@{}}T1w, \\ T2F, 1\%\end{tabular} & 0.64 & 0.58,   0.70  & 2.57 & 2.38,   2.76 \\ \hline
\end{tabular}
\caption{Performances for $\Delta$SUVR Mean and $\Delta$SUVR STD. All values are upscaled by 100 for better visualization purposes. The best performances in both tasks are in bold. For zero-dose task, the best performance is SGM-KD using T1w and T2F, while for ultralow-dose task, the best performance is TransUnet using T1w and PET 1\%.}
\label{tab:suvr}
\end{table}

Lastly, TransUnet demonstrates significantly faster sampling speeds compared to diffusion models, completing the entire volume in under 40 seconds, whereas SGM-KD requires approximately 400 seconds and SGM-VP takes around 2420 seconds.

\section{Discussion}
In this study, we comprehensively examined the performance of both TransUnet and score-based generative models to synthesize full dose FDG brain PET imaging in epilepsy patients using inputs that include either only MRI or MRI plus ultra-low 1\% dose PET images. We found that score-based diffusion models demonstrate stable performance, are less dependent on ultralow-dose PET inputs, and outperform a TransUnet model for zero-dose tasks. However, they suffer from slower sampling and more noticeable slice-to-slice intensity inconsistencies, which are suboptimal for volumetric evaluation. Adding T2 FLAIR MR images produces various effect depending on the model type. For instance, for the zero-dose task, adding T2 FLAIR improves the performance of SGM-KD in the zero-dose task, but slightly deteriorates the performance of SGM-VP and TransUnet. All of the models demonstrated ICC values suggesting at least good reliability.

Previously, Xiang et al. developed a CNN-based patch model for low-dose (25\%) PET denoising that incorporated T1w MRI information \cite{XIANG2017406}. Their approach used low-dose PET data acquired during a 3 min short scan extracted from a 12 min standard-dose tracer injection, which differs from the event under-sampling strategy we employed during list-mode reconstruction. Another prior study examining only MR-to-PET image modality synthesis integrated transformer, generator, and discriminator components into a U-Net architecture to synthesize FDG-PET images from multi-modal MR inputs, including T1w, T1c, T2 FLAIR, and ASL in brain tumor patients \cite{transunet}. While this approach demonstrated strong performance based on metabolic metrics such as SUVR, it relied on the availability of diverse contrast images, which may not always align with clinical acquisition protocols. In contrast, the current study utilizes fewer and more basic MRI inputs, improving the practicality and generalizability of the method across diverse datasets. Additionally, the availability of list-mode data provides higher flexibility to simulate different dose levels, facilitating future explorations into dose optimization. 

Due to the scarcity of publicly available PET datasets, particularly for epilepsy cases, we first examined prior work on PET reconstruction and synthesis in dementia studies. The availability of large ADNI datasets has enabled the development of multiple GAN-based 3D image translation approaches \cite{li2014deep, hinge2022zero, sikka2018mri}. For instance, Li et al. applied patch-based 3D CNNs for MRI-to-PET translation using 19.9 million patches from ADNI \cite{li2014deep}. Hinge et al. employed a Pix2Pix conditional GAN \cite{hinge2022zero}, while Sikka et al. utilized a 3D U-Net on full MR volumes to capture global context, improving Alzheimer's classification using synthetic PET \cite{sikka2018mri}. Similarly, Pan et al. used a cycle-consistent GAN to impute missing PET data, integrating it with MRI for AD diagnosis \cite{pan2018synthesizing}. With a focus on clinical applications, Choi et al. trained a variational autoencoder (VAE) on normal brain PET to assist in AD and MCI diagnosis \cite{choi2019deep}. For amyloid PET reconstruction, Ouyang et al. proposed a 2.5D GAN model with task-specific perceptual loss, leveraging a pre-trained amyloid status classifier to enhance pathological feature matching \cite{ouyang2019ultra}. Additionally, the multi-Pareto GAN introduced a multi-round generator and dynamic Pareto discriminator to iteratively denoise low-dose PET scans \cite{fu2024mpgan}.  Despite these advancements, the exploration of 3D diffusion architectures remains significantly underdeveloped, likely due to their high computational demands. Diffusion models hold great promise for real-world applications in image processing and acquisition. For example, a denoising diffusion null-space model was trained on high-quality PET images from a high-end scanner and integrated into a conventional PET system to enhance lower-quality scans \cite{zhang2024realization}. This highlights the realistic potential of diffusion-based methods to improve PET imaging, warranting further investigation into their feasibility and impact.

Specifically in the field of epilepsy, Yaakub et al. synthesized what they term “pseudo-normal” PET images from T1w MRI using a GAN trained on healthy subjects \cite{pseudo_normal_pet}. This model was then compared with an acquired FDG brain PET using a z-score metric, with areas of discrepancy highlighting potential seizure-affected regions, demonstrating about 92\% accuracy for localization of the epileptic zone. Unlike their emphasis on comparing a full-dose FDG brain PET image to a synthesized pseudonormal scan for the purposes of diagnostic insight, our work prioritizes reducing radiation exposure while emphasizing accurate clinical indicators such as hemisphere asymmetries.

There are several important considerations when choosing a generative model synthesizing full dose FDG PET. TransUnet, while having the fastest sampling ($<$40s per volume), needs larger training data or more input information for zero-dose task. Thus, it is the most suitable for larger training sets containing both MRI and low-dose PET images. Of the two score-based diffusion models, SGM-KD has shorter sampling time than SGM-VP due to its use of step discretization. Specifically, we can limit the number of sampling steps to 50-150 steps ($<$37s per 8-slice inference batch, ~400s per volume) for a satisfactory result, compared to 200-300 steps that are required for the SGM-VP ($<$220s per 8-slice inference batch, ~2420s per volume). Further exploration on how to reduce the sampling time of SGMs are ongoing, including consistency models \cite{consistency_model} and DPM samplers \cite{dpmsolver}. 

The best model also depends differently on the evaluation metric of interest and the available inputs. SGM-KD, while having the least mean $\Delta$SUVR and the highest ICC for zero-dose task, may not be as accurate as SGM-VP in terms of capturing hemisphere asymmetries. For the ultralow-dose task, TransUnet outperformed in congruency measures, $\Delta$SUVR measures, and sampling efficiency, but has slightly lower performance in ICC, though still in the range indicating good consistency. This changes the image translation task to an image denoising task, emphasizing the potential in developing better ultralow-dose image reconstruction techniques.

It is important to note that the limited training data and variations in model architecture and training time may contribute to discrepancies in model performance. These differences may diminish with access to larger datasets, as models typically converge to better performance when trained on more comprehensive data. Our study relied on single-center data, minimizing harmonization challenges but potentially oversimplifying the task compared to using multi-center datasets. The ultralow-dose PET was simulated from full-dose injections, which may not fully represent PET images acquired with lower dose levels, though prior studies with true ultralow dose have suggested good concordance with similar simulations \cite{chen2021ultra}. Additionally, our analysis was restricted to a subset of deep learning models commonly applied in medical imaging, leaving room for broader exploration. Finally, a reader study to evaluate the clinical utility of the synthetic PET images remains an important area for future investigation.

\section{Conclusion}
In an epilepsy patient population, we demonstrate the ability to synthesize full-dose FDG-PET brain images from MRI inputs using deep learning diffusion models and the potential to improve performance by including ultralow 1\% dose FDG-PET images as inputs. Future work will focus on refining slice consistency through averaging techniques \cite{choo2024}, exploring 3D full volume and patch-based approaches, improving ultra-lowdose reconstruction quality, and assessing whether any of the evaluation metrics such as Congruence Index (CI) correlates with findings from reader studies. Additionally, we plan to explore multi-task brain MRI-PET translation, including different tracers (amyloid, tau) and different disease conditions (tumor, dementia, etc.). Incorporating clinical reports as additional conditioning inputs may further enhance model performance. Overall, this study highlights the potential of generative AI approaches to significantly reduce FDG-PET dose levels, offering substantial clinical benefits for patients.

.




\bibliography{sn-bibliography}

\end{document}